\documentclass[a4paper]{spie}  
\usepackage[]{graphicx}

\title{A Fiber Positioner Robot for the Gran Telescopio Canarias}

\author{M. Azzaro\supit{1}, S. Becerril\supit{1}, C. Vilar\supit{3}, X. Arrillaga\supit{2},
        J. S\'anchez\supit{1}, I. Morales\supit{1}, M. A. Carrera\supit{2} and F. Prada\supit{1}
\skiplinehalf
\supit{1}Instituto de Astrof\' isica de Andaluc\' ia (IAA-CSIC), C. Bajo de Hu\'etor n.50, 18008 Granada, Spain; \\
\supit{2}Added Value Solutions, Eibar, Guipuzkoa, Spain; \\
\supit{3}Universitat de Barcelona, Barcelona, Spain
}

\authorinfo{Further author information: (Send correspondence to M. Azzaro at mazzaro@iaa.es,
            or to S. Becerril at becerril@iaa.es)}

\begin{document} 
  \maketitle 

\begin{abstract}
Fiber-fed spectrographs dedicated to observing massive portions of the sky are increasingly
being more demanded within the astronomical community. For all the fiber-fed instruments,
the primordial and common problem is the positioning of the fiber ends, which must match
the position of the objects of a target field on the sky. Amongst the different approaches
found in the state of the art, actuator arrays are one of the best. Indeed, an actuator array
is able to position all the fiber heads simultaneously, thus making the reconfiguration time
extremely short and the instrument efficiency high.
The SIDE group\footnote{see http://side.iaa.es} at the Instituto de Astrof\'\i sica de
Andaluc\'\i a, together with the
industrial company AVS and the University of Barcelona, has been developing an actuator
suitable for a large and scalable array. A real-scale prototype has been built and tested
in order to validate its innovative design concept,
as well as to verify the fulfillment of the mechanical requirements.
The present article describes both the concept design and the test procedures
and conditions. The main results are shown and a full justification
of the validity of the proposed concept is provided.
\end{abstract}


%
%
%
%
%
\section{Introduction}
\label{sec:intro}  
Extremely important results in observational astrophysics are obtained
today through large databases of spectra or images (e.g. SDSS, 2dFGRS). In order to allow
this statistical approach to science, survey instruments are becoming
more and more necessary. Such instruments are dedicated to observing
massive portions of the sky, and must be as efficient as possible in
order to minimize times, therefore the largest possible number of
objects must be observed at the same time.
Concerning Spectroscopy, there are two main types of Multi-Object
instruments: multi-slit spectrographs and fiber-fed spectrographs.
Both offer advantages and disadvantages, but
the most versatile type for object collection is the fiber-fed device.
For all the fiber-fed instruments, the primordial and common problem is
the positioning of the fiber ends, which must match the position 
of the objects of a target field on the sky. For each field to be observed,
the configuration of the fibers is different, so, approximately every hour
of an observing night, all the positions of the fibers must be changed.
Unless a dedicated person takes care of it (as for the SDSS), there are
two broad groups of devices (usually called robots) for this task:
pick and place devices or actuator arrays moving one fiber head each
(see Smith et al. \cite{smith} for a general review, or Haynes et al. \cite{haynes}
for a review on actuator technology).

A pick and place device has one moving gripper which grabs one fiber head
from its parking position and places it in position, usually on a magnetic
plate; it can move one fiber head at a time, thus reconfiguration time
scales with the number of fiber heads. This reconfiguration time is
essentially time lost from observation (e.g. Autofib at WHT, Flames at VLT).
An actuator array is able to position all the fiber heads
simultaneously, making the reconfiguration time extremely short and the
instrument efficiency high (e.g. LAMOST, Echidna)

The SIDE group at IAA, together with the industrial company
AVS and the University of Barcelona, has been developing an actuator suitable
for a large and scalable array. The mechanical design presented here is the
result of a development based on an idea employed  in the LAMOST project
(see Zhang et al. \cite{lamost}), improved by the LBNL in Berkeley
(see Schlegel et al. \cite{lbnlrobot}) and
finally developed into a substantially different design and a prototype.
The present article describes both the concept design and the
test procedures and conditions. The main results are shown and a full justification
of the validity of the proposed concept is provided.

%
%
%
%
%
\section{The Focal Plane Array}
\label{sec:FocPlane}  
\begin{figure}
   \begin{center}
   \includegraphics[width=10cm]{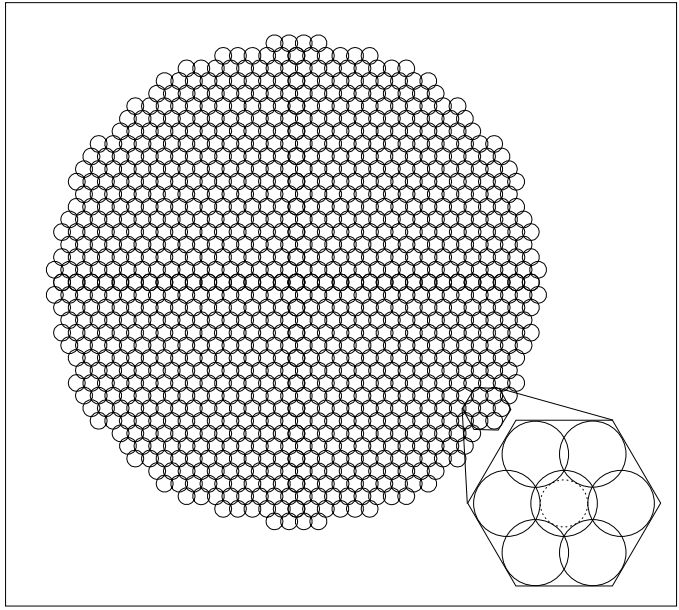}
   \end{center}
      \caption{The array of cells into which the focal plane is divided;
               each actuator of the robot is dedicated to one of them, thus each fiber
               head covers one of these cells. The so-called Patrol Discs overlap
               so that all the focal plane is covered. This example features 1003
               cells for the 992 mm diameter field of view at GTC. The detail shows a group of
               7 Patrol Discs and how they overlap, as well as the Security Circle (dotted),
               which is the zone where the fiberhead cannot be touched by other actuators.
              }
         \label{FocPlaneActuators}
   \end{figure}

This section is devoted to describing the robot as a
complete set of actuators which fills the focal plane of the
telescope.
The concept used is a single actuator for each fiber head to be
positioned, then the focal plane is populated with an array of such
actuators, distributed with hexagonal pattern, each devoted to
observing a single object.
In a few words, this Fiber Positioner is a collection of
actuators, all identical, distributed over an array which covers the
focal plane. Each actuator can position a fiber head over a disc which
it will be called {\it Patrol Disc}. The focal plane is then covered
by these Patrol Discs so that all positions can be reached by at least
one actuator (see Fig.~\ref{FocPlaneActuators}).
Of course, some parts of the focal plane can be reached
by more than one actuator, hence the possibility of collisions.
   \begin{figure}
   \begin{center}
   \includegraphics[width=10cm]{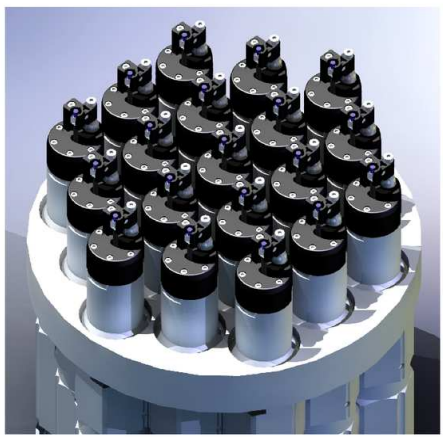}
   \end{center}
      \caption{A view of a subset of the actuator array mounted on their holder
               (here, 19 units are shown as an example).
              }
         \label{19_actuators}
   \end{figure}
Fig.~\ref{19_actuators} shows a view of a sample subset of the complete array,
with $19$ actuators in hexagonal pattern.\\
This concept offers many advantages with respect to others. Concerning
the scientific advantages, reconfiguration times are extremely short
(thus observation overheads basically depend on detector readout times
and calibrations). In addition the differential atmospheric dispersion (see
Donnelly et al. \cite{donnelly}) can be corrected in real time, thus
a very large field of view can be used more efficiently. Such correction
is called {\it ADT} (Atmospheric Differential Tracking) throughout this article,
referring not to a telescope tracking but to the differential atmospheric
dispersion real-time compensation of the affected actuators.
Concerning the mechanics or hardware, the difficulties which can arise from the
need of placing the actuators on a curved (spherical) field of view are further
detailed in the next section.
This concept is extremely robust, scalable and easy to
service and maintain (failure of one actuator causes the loss of one
object only).
There is a drawback concerning the science: the actuators cannot be densely packed
onto a small portion of the field of view, so the system is efficient for rather
uniform distributions of objects. This deficiency is compensated by the time
efficiency of this design.
Concerning the control and software, an efficient algorithm must be
developed to control 2000 motors, while avoiding collisions between
fiber heads.

Although our group has faced the Fiber Positioner from a broad point of
view (holder, focal plane topology, collision problems), this article will
concentrate on a single actuator as an example of array unit, for which we
built and tested a prototype.

%
%
%
%
%
\section{Applicable boundary conditions}
\label{sec:Boundcond}  
\begin{figure}
   \begin{center}
   \includegraphics[width=10cm]{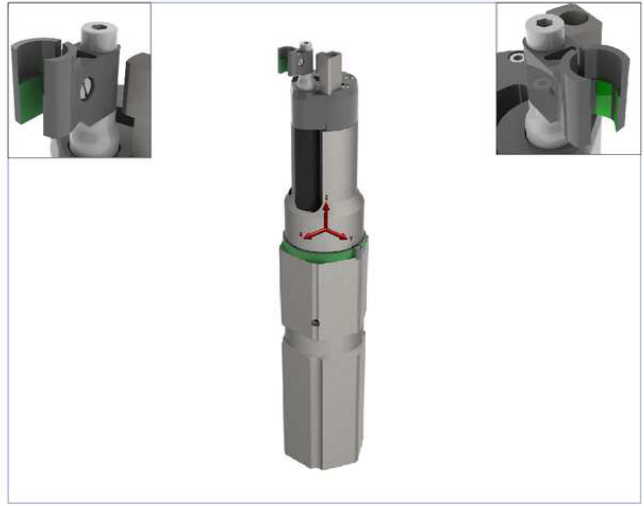}
   \end{center}
      \caption{A view of the applicable interfaces of the actuator prototype.
               The small panels (left and right) show the interfaces with the
               Fiber Button (coloured green), and the central panel shows
               the interface with the actuator's array holder (coloured green).
              }
         \label{interfaces}
   \end{figure}

The boundary conditions from which the main geometrical and envelope requirements
are defined concern the telescope focus where such device would be mounted.
In practical terms, the Nasmyth Focus at the GTC Telescope has
been used as a reference. Indeed, such reference sets a reasonable framework
within the potential instrumentation for 8-10m telescope. Thus, the actuator was
originally designed for a $992\,mm$ field of view (equivalent to $20\,arcmin$) and a
density of objects of $3.19 \,Obj \cdot arcmin^{-2}$, which set the number of
actuators (1003 units). The applicable Focal Plane is a concave spherical cap with
$3574\,mm$ of radius of curvature.
The hexagonal pattern packing has been chosen for the array of actuators since it
provides the densest population. Thus, the centre-to-centre distance
between actuators derived from the above constraints is $29.2 mm$.
Other important factors to be known about the telescope are the (foreseen) pointing
error ($0.1\,\,arcsec$), the plate scale ($825\,\mu/arcsec$) and the focal length
set at F/15.
Another boundary condition comes from the fact that the original fiber head
was formed by a honeycomb-shaped 7-microlens array, which focused the light
into 7 fibers bundled together. The encircle diameter of the 7-microlens
array is 1.2mm.
The actuator is limited, on one side, by the interface of the chassis with
the Focal Plane Array holder and, on the other side, by the mechanical
references of the Fibers Button (see Fig.~\ref{interfaces}).\\
The actuator coordinate system used through this work was defined as follows:
Z is the optical axis (the axis along which the light travels)
and the plane XY is orthogonal to it. As a result, a displacement along Z
produces de-focus while a displacement on the plane XY produces de-center of
the image.
It is also relevant the angle between the Fiber Button axis and Z, which
reverts directly on the budget of the angular error of the fiber with respect
to the optical axis.
The manufacturing and mounting errors of
the fibers bundle and microlenses on the Fibers Button are beyond the scope
of the present work.
%
%
%
%
%
\section{Mechanical requirements}
\label{sec:MechRequirem}  
The mechanical requirements of the Fiber Positioner are shown in Table~\ref{ActuatorRequirements}.
   \begin{table}
      \caption{Actuator basic requirements}
         \label{ActuatorRequirements}
         \begin{center}
         \begin{tabular}{c c}
            \hline
            \rule[-1ex]{0pt}{3.5ex}  Item & Value \\
            \noalign{\smallskip}
            \hline
            \noalign{\smallskip}
            \rule[-1ex]{0pt}{3.5ex}  Packing Pattern Geometry & Hexagonal \\
            \rule[-1ex]{0pt}{3.5ex}  Distance between actuator centres & 29.2 mm \\
            \rule[-1ex]{0pt}{3.5ex}  X/Y Position Repeatability & $40\,\mu$ radius\\
            \rule[-1ex]{0pt}{3.5ex}  X/Y Position Accuracy & $100\,\mu$ radius\\
            \rule[-1ex]{0pt}{3.5ex}  Z max defocus error & $\pm 140\,\mu$ \\
            \rule[-1ex]{0pt}{3.5ex}  Max angular tilt & $\pm 1 \,mrad$ \\
            \rule[-1ex]{0pt}{3.5ex}  Reconfiguration time & $60\,sec$ \\
            \rule[-1ex]{0pt}{3.5ex}  Working temperature & $-10 \,\, +30 \,{}^{\circ}$C \\
            \rule[-1ex]{0pt}{3.5ex}  Weight of the array  & $750 \,Kg$ (Holder not included)\\
            \rule[-1ex]{0pt}{3.5ex}  Lifetime  & $10$ years\\
            \noalign{\smallskip}
            \hline
         \end{tabular}
         \end{center}
   \end{table}
In order to understand the origin of some of the requirements, it is
necessary to define the errors which affect one actuator and distinguish
between the errors for which subsequent software compensation is feasible and those
for which it is not. It is clear that the requirements shown in Table~\ref{ActuatorRequirements}
must be fulfilled by the actuator affected by these last errors (because it
is too expensive to compensate them).
Sources of XY errors are the following:

\begin{itemize}
\item[1.] XY De-center of the actuators housings machined in the Focal Plane Array
Holder: Here the shape of the Focal Plane has heavy consequences in terms of
error budget. According to the boundary conditions mentioned above, the Focal
plane is a spherical cap, which implies that all the actuators housing on the
Focal Plane Array must point (within certain tolerance) to the centre of curvature of the
Focal Plane. From the manufacturing point of view, this presents a much higher degree of
complexity than the case of a flat focal plane and errors are more likely.
\item[2.] Dimensional errors of the actuator affecting the position of the
fiberhead on its Patrol Disc: indeed, the manufacturing and assembly errors applied to
the mechanical chain from the chassis to the part which holds the fiber
bundle may lead to a decenter of the Patrol Disc with respect to its theoretical
location. This would add to the fiberhead position errors due to the actuator itself.
\item[3.] Gearing errors: due to machining errors in the commercial gearboxes used
in the actuator, noticeable XY errors may result in the final position of the fiberhead.
\end{itemize}

While error 1 above is part of the array holder budget and its characterization
involves accurate measurement on one part only (the holder),
errors 2 and 3 individually affect each of the 1003 actuators of the
array, thus any additional process needed to mitigate these would revert in
extremely high costs for the device.
It is therefore assumed here that only error 1 could be characterized and
compensated by software, once the Fiber Positioner is assembled, and so the
requirements shown in Table~\ref{ActuatorRequirements} must be
fulfilled by the actuator affected by errors 2 and 3.

Concerning de-center, the maximum error has been taken as $10\%$ of the
incircle diameter of the microlenses set, which yields $120\,\mu$.
However, this figure includes the errors of microlenses and fibers manufacture
and mounting, which have been budgeted at $66\,\mu$
(according to the state-of-art fiber optics technology). As a
result, $100\,\mu$ radius is the value applicable to the actuator.

Concerning the tilt angle of the fiberhead, the main driver is the
admissible angular deviation of the light beam entering the
fibres: $2\,mrad$. This error is not compensable and it adds to tilt errors
of both the Focal Plane Array holder and the fiberhead/microlens
assembly. Once these last ones are subtracted, the value of $1\,mrad$ is
left for the actuator.

Concerning the defocus, optical analysis shows that a de-focus
error of $400\,\mu$ would imply only $1\%$ of energy losses. This is
due to the large focal length of GTC. Once again, this error is not compensable.
The main budgets here to be taken into account come from the actuator and the Focal Plane Array
Holder. $350\,\mu$ have been conservatively assigned to the holder
due to the difficulties of machining a spherical surface.
Therefore, $140\,\mu$ are left and applicable to the actuator.

Concerning the XY precision, it must fit with the pointing error of the
telescope ($0.1\,arcsec$), which means $40\,\mu$ radius at the Focal Plane.

According to the GTC requirements, the Nasmyth Rotator can withstand $2400\,kg$
with no auxiliary support. This figure includes $1003$ actuators, the Focal Plane
Array holder and the mechanical structure for attachment to the Nasmyth
Rotator. The weight budget for the $1003$ actuators only is $750\,kg$.

Finally, the reconfiguration time of the actuator array must be comparable
to the time spent in reading out the detector and data archiving, in order to
minimize overheads. Thus, $60$ sec has been set. 

%
%
%
%
%
\section{The actuator concept design}
\label{sec:ConceptDesign}  
\subsection{The mechanical design}
\begin{figure}
   \begin{center}
   \includegraphics[width=10cm]{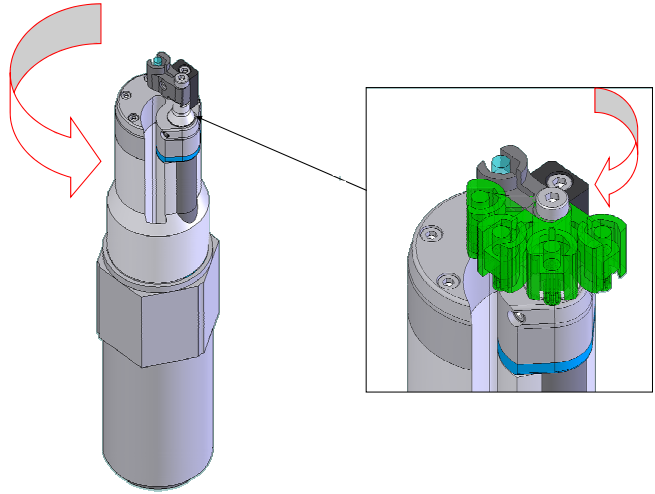}
   \end{center}
      \caption{A view of the assembled actuator prototype (left), with its
               hexagonal steady frame and the rotating internal cylinder.
               The other view (right) is a zoom-into the R2 rotation.
               The arrows show the direction to reach the zero stop of
               each axis. The slot cut into the main rotating cylinder
               to accommodate the fibers is also visible.
              }
         \label{ROT1ROT2}
   \end{figure}

\begin{figure}
   \begin{center}
   \includegraphics[width=10cm]{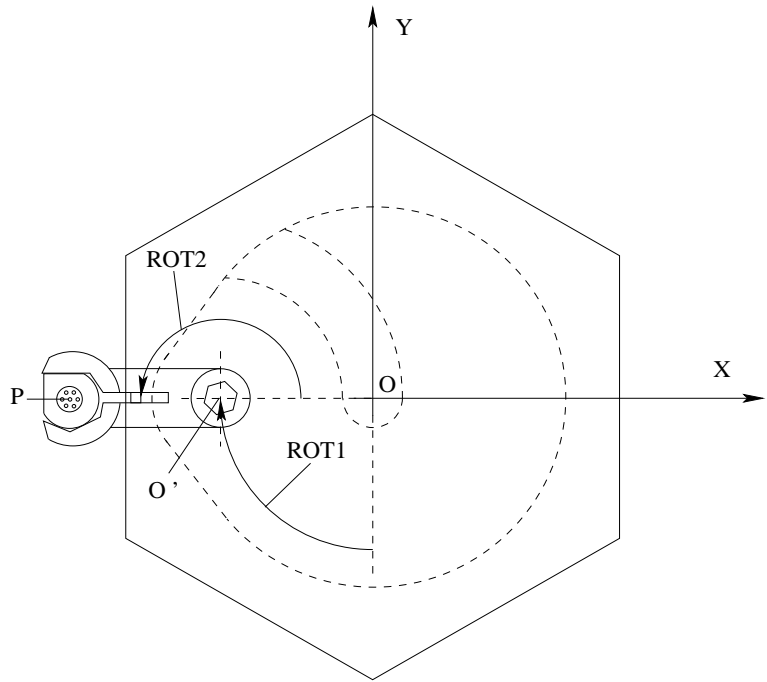}
   \end{center}
      \caption{The rotation directions of the actuator. P is the fiber
               head centre, O is the centre of R1 and O' is the centre of R2.
              }
         \label{RotDir}
   \end{figure}

The mechanical design presented here is the result of a development based
on an idea employed in the LAMOST project, improved by the LBNL in Berkeley;
the Cobra actuator (see Fisher et al. \cite{cobra}) also has many similarities
with this design.
The basic design of the actuator was developed by the company AVS in collaboration
with the IAA-CSIC and can be seen in Fig.~\ref{ROT1ROT2}.
This design is substantially different from both the LAMOST and the LBNL
designs; we believe that, mainly, robustness, reliability and simplicity are
improved with respect to these.
The fiber head is brought around through two rotations in cascade: the
first rotation, called ROT1 (or R1), involves the internal cylinder of the
actuator (in which the motor for the second rotation is embedded), and
takes place with respect to the hexagonal frame which should be attached
to the actuators' holder. The second rotation, called ROT2 (or R2),
moves an arm which ends into a clamp where the fiber head is held.
This movement takes place with respect to the ROT1 rotating cylinder.
A slot (visible in Fig.~\ref{ROT1ROT2}) is cut into the main cylinder,
in order to accommodate the fiber
which runs from the fiber head to the back of the actuator. The fiber is
both twisted and bent, but this takes place over a length of about 25 cm
and the fiber is protected by a plastic pipe. The electronics board is planned
to sit at the back of the actuator, attached to the internal rotating cylinder,
so that less cables need to be twisted (only the power and signal for the board,
instead of the cables for the two motors). Counter-posed directions minimize the
total twisting of the fibers/cables with respect to their rest position.
Fig.~\ref{RotDir} shows how the rotation directions were chosen.
There are physical stops which limit the range of the two rotations,
so that the optical fibers and cables cannot be twisted beyond a safe amount.
The stops allow about $365^{\circ}$ for R1 and about $230^{\circ}$ for R2. 
One step motor is used for each axis, with a reducer gearbox and an
incremental encoder. For convenience, one motor is run at full steps (R1), while
the other runs at half steps (R2). A design review (PDR) by AAO and LBNL engineers
took place in late 2008 and many useful suggestions were given which would further
improve the final actuator.

\subsection{The control electronics}
\begin{figure}
   \begin{center}
   \includegraphics[width=10cm]{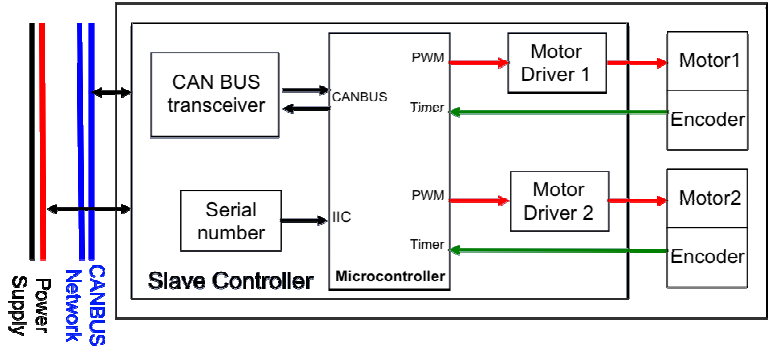}
   \end{center}
      \caption{Scheme of the actuator controller.
              }
         \label{electronics1}
   \end{figure}

The Fiber positioner robot electronics has two main tasks: 
\begin{itemize}
\item[-] Control the actuator arm (slave controller)
\item[-] Communicate with the master controller
\end{itemize}
Each actuator has a dedicated control electronics (see Fig.~\ref{electronics1})
fixed at the actuator rear part. This electronics provides a mean to solve both
tasks with a limited size and power budget.\\

{\bf Control of the actuator arm.}
The actuator controller (Fig.~\ref{electronics1}) has the following requirements:
\begin{itemize}
\item[1.] Control the two step motors in a closed loop operation mode in order to
move the actuator arm with the required precision and reconfiguration time.
\item[2.] Implement a safety low-level protocol to avoid, detect and report to the
Master controller any collisions between neighbouring robot arms.
\item[3.] Perform an actuator logical zero calibration without using an electronic
limit switch detector.
\end{itemize}
The actuator controller is based in a 32 bits Coldfire V1 micro-controller
with an extended timer operation functions in order to control the 2
steppers motors and read the encoders information. It is important to
prevent any heat source inside the telescope focal plane, thus the
micro-controller and all electronic devices must be low power in the active
(robot positioning period) and sleep modes (observation period).
The motors used in the prototype were 8mm diameter stepper motors, manufactured
by Faulhaber, each moving one axis in combination with a 1:120 gearbox and
a quadrature encoder with 32 lines resolution.
This motor and gear selection produces enough torque to move the fiber
button and enough holding torque to hold the science fiber position securely. 
It is interesting to note the difficulty to calibrate the actuator logical
zero by using a spring buffer due to the elastic stop of the gearbox.
It has been necessary to implement a procedure to recover a
non-elastic step movement through the encoder measurement.\\
   \begin{figure}
   \begin{center}
   \includegraphics[width=10cm]{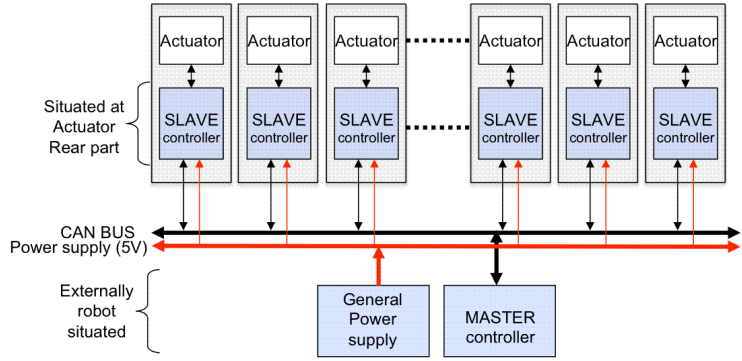}
   \end{center}
      \caption{Scheme of the Fiber positioner network.
              }
         \label{electronics2}
   \end{figure}

{\bf Communication with the master controller.}
All the $\sim 1000$ robot actuators are controlled by a Master computer thought a CAN
Bus 2. A network (see Fig.~\ref{electronics2}) in a Master-Slave operation mode.
Each actuator has a CANbus transceiver with a CAN address provided by a serial number
circuit in order to be identified. 
The CAN transceiver used in the actuator controller can not drive the $\sim 1000$
actuator network. Due to the transceiver fan-out limitation, it is necessary
to use a network switch to interconnect all actuators in a tree topology.

%
%
%
%
%
\section{Scope of the actuator prototype}
\label{sec:ScopeofProto}  
A prototype of the actuator was manufactured in order to submit it to a
comprehensive testing plan aimed to check whether the mentioned requirements
can be fulfilled by this design.

\begin{figure}
   \begin{center}
   \includegraphics[width=8cm]{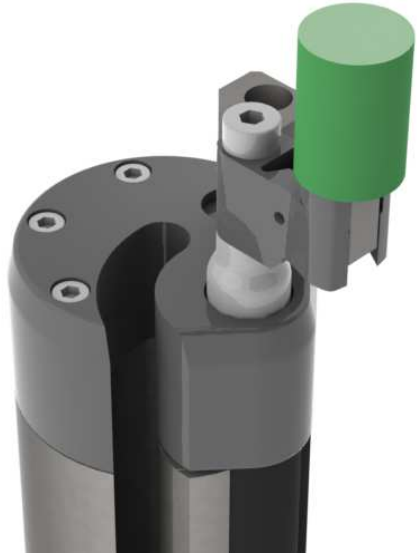}
   \end{center}
      \caption{A view of the test button (green) held in the
               prototype clamp for the fiberhead.
              }
         \label{testbutton}
   \end{figure}

From the mechanical point of view, the prototype is an exact copy of an
actuator. Therefore, if the prototype fulfils the requirements,
a real actuator also does. The main
differences lie on the button and the control electronics.
Indeed, a special button (Test Button) has been designed and
manufactured in order to improve the measurement procedures
(see Fig.~\ref{testbutton}). This button presents appropriate surfaces
for being measured by the 3D coordinate measuring machine.

\begin{figure}
   \begin{center}
   \includegraphics[width=10cm]{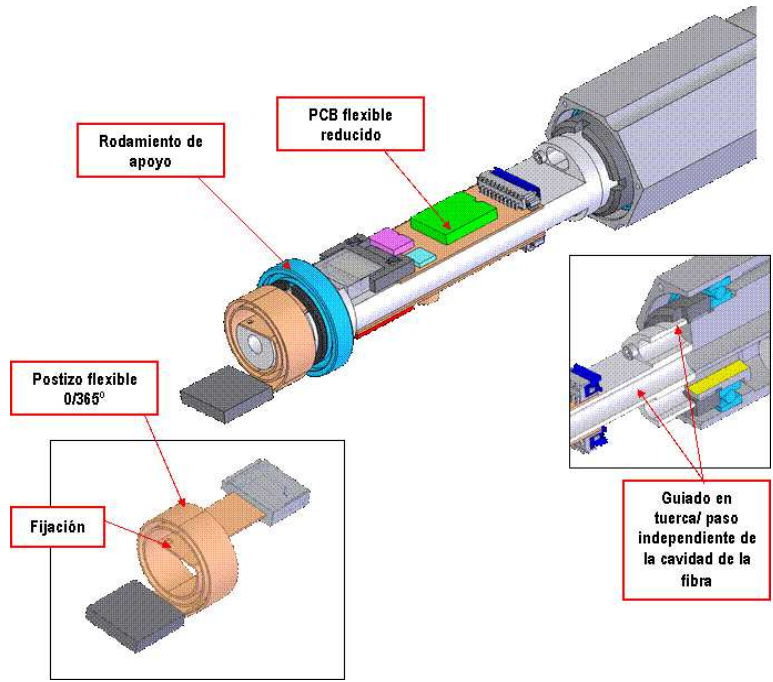}
   \end{center}
      \caption{A view of the real control electronics foreseen for one
               actuator. The central panel shows a general view of
               the electronics attached to the back of the actuator.
               The bottom-left panel shows a detail of the flexible
               contact to connect the rotating part to the fixed
               structure. The right panel shows a detail of the
               central hole for the fiber optics.
              }
         \label{pcb}
   \end{figure}

Concerning the control electronics used in the prototype, this was
different from the one planned for a real actuator.
A rigid breadboard electronics has been used for the tests, instead of the
encapsulated flexible PCB (see Fig.~\ref{pcb}) foreseen in the 
actuator design. No major consequences derived from this
issue: the encapsulated PCB has negligible influence on the mechanical
performance (the real PCB is connected to the rotating part of the
actuator through a rolled flat contact part).

\begin{figure}
   \begin{center}
   \includegraphics[width=10cm]{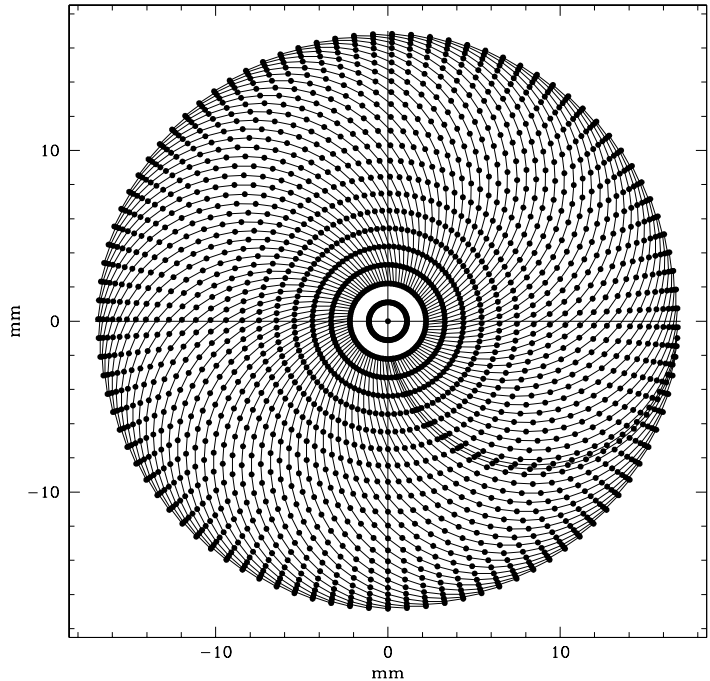}
   \end{center}
      \caption{Example (with much reduced number of points) of the actuator's grid
               of reachable positions. The denser zone in the bottom-right part of
               the picture is due to the superposition of points because of the
               extra run of 5 degrees of R1 beyond a whole turn. The multiple
               semicircles over the whole Patrol Disc are simply the path of
               the fiber head when moved in R2.
              }
         \label{GridEx}
   \end{figure}

Finally, a remark is due here to the fact that the performance presented is
relative to the discrete grid of positions (Fig.~\ref{GridEx}) of the Patrol Disc,
defined by the finite resolution of the step motors.
In other words, only positions belonging to this grid have been submitted to test.
In general, the position of an object rarely matches a point of the grid
(composed by more than 23 millon reachable positions) which implies an extra
source of error which is not within the scope of the present work.
Anyway, the resolution available in ROT1 and ROT2 limits this error
to 5.5 microns in the less favourable cases (outer areas of the Patrol Disc).

%
%
The resulting grid of positions is such that any random position on the Patrol Disc
can be approximated by a grid point with no more than $5.5 \mu$ of error.

When several actuators are mounted in array on focal plane, their Patrol Discs
overlap and collisions amongst fiber heads are possible.
In particular, collisions are possible for a fiber head when it is outside a
circle which we call ``Security Circle'' (see detail of Fig.~\ref{FocPlaneActuators}).
This risk can be canceled by a clever control software.

%
%
%
%
%
\section{Testing procedures}
\label{sec:TestProc}  
The testing of the prototype has been planned by means
of two different experimental setups: an optical setup for
the measurements where high X-Y resolution over a tiny FoV was sufficient,
and a mechanical setup (based on a 3D coordinate measuring machine) for
the measurements where high precision over a measuring range of several
millimeters across the Patrol Disc was required.
This dual approach on the testing method was imposed by
logistics reasons: the 3D coordinate measuring machine had very limited
availability to our group, thus the need of the optical setup.
Also, when dealing with a prototype, high flexibility during the tests is
very important, so the comprehensive set of repeatability tests has
been implemented in-house at the IAA through the optical setup, while
the tests about position accuracy are planned to take place at IMH
(Instituto de M\'aquina-Herramienta, Elg\'oibar, Spain).
This set of measurements is much smaller than that of the repeatability tests
and, at this stage, is still pending, therefore only the optical setup is
detailed here.\\
One drawback of the optical setup was that no tip/tilt or defocus could
be measured.

As stated previously, we will use the convention that the plane X-Y contains
the Patrol Disc, and the Z coordinate is parallel to the actuator's long axis,
positive from the back to the front of the actuator.

\subsection{Optical setup}
\begin{figure}
   \begin{center}
   \includegraphics[width=10cm]{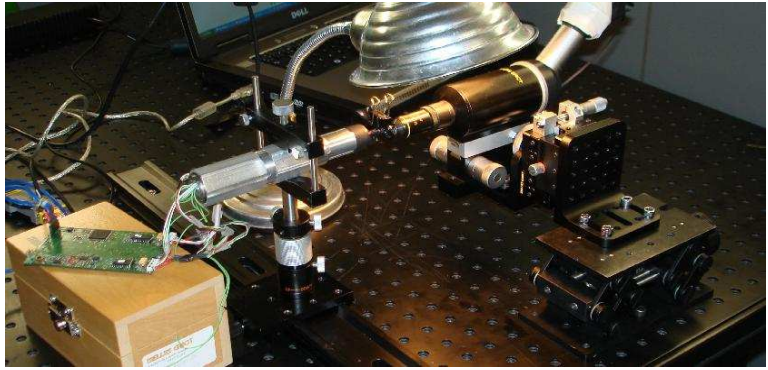}
   \end{center}
      \caption{A view of the optical setup used for some of the tests:
               on the optical bench, a support holds the actuator prototype
               (left), and another support holds the microscope used to
               inspect the actuator movements.
              }
         \label{optsetup}
   \end{figure}

\begin{figure}
   \begin{center}
   \includegraphics[width=10cm]{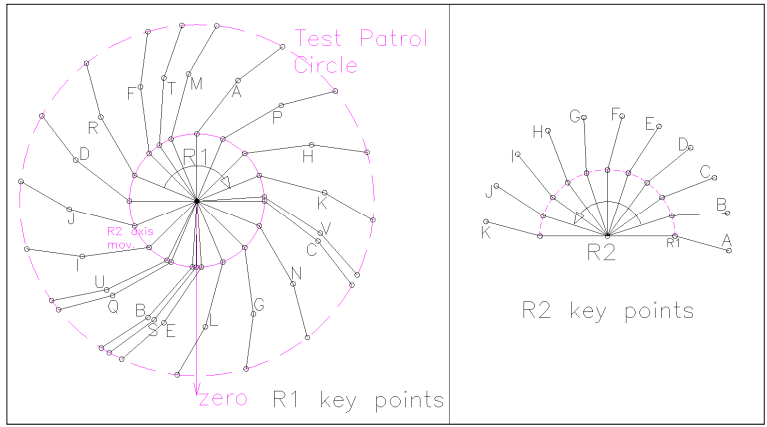}
   \end{center}
      \caption{A sketch of the distribution of the Test points
               over the Patrol Disc.
              }
         \label{testpoints}
   \end{figure}

This setup is based on a microscope with a hi-res camera attached (see Fig.~\ref{optsetup}).
Both the microscope and the prototype are mounted on multiple-stage
sliding supports, thus providing the capacity of measuring at any point
of the Patrol Disc. A $2\,\mu$-size pinhole is attached to the
Test Button of the prototype. Both supports are mounted on a rigid steel
plate which, in turn, is kinetically held by means of three screws
on the test bench. In addition, a cold lamp is used to lighten the setup.
This way, the light coming through the pinhole is captured by the camera
attached to the microscope. By connecting the camera to a laptop computer,
it is easy to capture and store images of the spot (projected pinhole).
The magnification of this optical setup allows capturing a FoV of about
$780 \times 580\,\mu$ on a picture of $640 \times 480$ pixels, resulting
into a scale of $1.22\,\,\mu/pix$.
Due to the relaxed requirement in Z ($400\,\mu$), we measured only X-Y
repeatability, with confidence that the behaviour in Z was far below
the corresponding requirement.
The tests so implemented have produced a comprehensive series
of pictures, whose spot positions have been analyzed through the image-processing
package IRAF and, finally, by statistics routines.
Finally, in order to ensure that any thermal drift is avoided
over the duration of a single test, an air fan is placed next to the
measuring area.
Fig.~\ref{testpoints} shows the layout of the key points (the centres of both
rotations, the centre of the Test Button and the pinhole) involved in the
optical tests, both for ROT1 and ROT2. Note that the item which is physically
measured by the optical setup is the pinhole spot, whose location lies
beyond the Patrol Disc. Therefore, the data process must include a corrector
factor in order to translate the magnitudes measured at the pinhole to the
nominal radius. This has the advantage of increasing the accuracy of the measurements.
For ROT1, the nominal radius is that of the Patrol Disc, i.e.
$16.86\,mm$, with ROT2 set to the maximal extension ($2400$ half-steps nominally).
For ROT2, the nominal radius is the distance between the centre of ROT2 and
the centre of the Fiber Button, $8.43\,mm$.
\section{Performance and test results}
\label{sec:Performance}  
The tests which have been implemented for a full mechanical verification of
the prototype are explained in detail through this section.

\subsection{Stepper motors performance}
\subsubsection{Stepping reliability}
\begin{figure}
   \begin{center}
   \includegraphics[width=8cm]{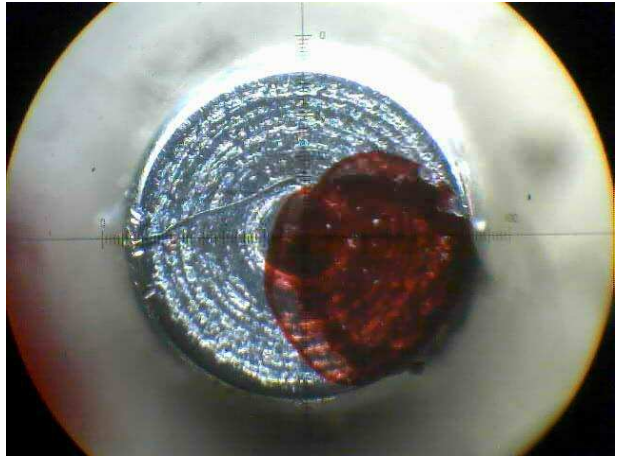}
   \end{center}
      \caption{A magnified view of the axis of a test motor.
               The bright spot on the right border of the axis served as
               a reference for the step angles calculations.
              }
         \label{motoraxis}
   \end{figure}

For the present work, the electronics was not yet developed for reading
the encoders of the motors. However, the reliability of the
electronics about the number of steps performed by the motors had to be
proved. This test was implemented on the motor+gearbox alone (not integrated
in the prototype) in order to avoid any effect from the rest of the mechanism.
It simply consisted of addressing several angular positions to the motors.
The back axis of the motor was marked so that a reference could be visually
checked. Since a motor turn implies 20 steps, the positions addressed were
always an integer multiple of 20. Thus, the mark at the back of the
motor axis should always stop at the same place. A camera viewing the back
of the motor allowed a detailed visual check (see Fig.~\ref{motoraxis}).
%

\subsubsection{Energy mode for the motors}
A parameter that was found important for the
motors performance is the energy mode which is set once a certain position
is reached. Indeed, the current to the motor can be configured as to be kept either
``disabled'' or ``enabled'' once the movement has finished. Therefore, this
test consisted of running series of 20 half-steps (one by one) at
different domains of the angular range of ROT2, both with ``disable'' and
``enable'' configurations. As a result, the configuration leading to a
better performance in terms of homogeneity about the amplitude of the steps
was the selected one for the implementation of further tests. The
measurements were only implemented on ROT2, but the conclusions so obtained
were also applied to ROT1 since both rotations are driven by identical
motors. When the current was ``disabled'', the RMS relative to the mean step
was found to be $\sigma = 5.95\%$; when the current was ``enabled'',
the RMS relative to the mean step was found to be $\sigma = 3.38\%$. The
``enabled'' mode was then programmed by default into the low level software
of the control electronics.

\subsection{Re-configuration time performance}
The re-configuration time has been estimated through a software simulation of a
worst-case event. The simulation reproduces the situation of several
neighbouring actuators and the time needed to disentangle a situation
of conflict, i.e. when collision amongst them would occur with straight movements.
The simulated worst-case event also involved the longest runs of the two axes of
one actuator, from a start position to a target position across the Patrol Disc.
In such case, the re-configuration time (with axes speed of $300\,full$-$step\,sec^{-1}$
in ROT1 and $300\,half$-$step\,sec^{-1}$ in ROT2) was about 65 seconds.\\
This slightly exceed the requirement value, but it must be noted that the
estimation is extremely conservative, as it is based on a worst-case that
software simulations prove to be very unlikely.\\
Also, if necessary, the axis speed of the motors could probably be increased.\\
It can be stated that, in the great majority of cases, the re-configuration
time requirement is met, and the extra delay added by the few worst cases
is negligible.

\subsection{Precision tests}
These measurements have been implemented by means of the optical setup
mentioned above. 
\subsubsection{Approach mode}
Although the gearboxes here selected are labeled ``zero-backlash'' by
the supplier, this test was performed in order to check whether it made
any difference to approach a certain angular position from one side or
the other. Strictly, a zero-backlash transmission should
not behave differently when changing the direction from which a position
is approached. By the way, this test was also useful to check whether, in spite
of its preloaded floating mount, the extra 1:4 gearing stage of ROT1 included
some backlash.
The test consisted of approaching different target positions, each of them
from two start positions. For a certain target position, each start position
was located at a different side as regard to the target. For the movements
from the start positions to the targets, two options were studied.\\
{\bf Single Turn mode.}
The movement from a start position to the target is made in a single movement.
This means that the rotation direction to the target position changes when
the starting position is on one side or the other of the target position.\\
{\bf Single Direction mode.}
the target position is approached always from the same direction, and such
direction was defined clockwise for ROT1 and counter-clockwise for ROT2.
This means that, if the start point is located between the zero and the
target, the movement equals the single turn mode.
Otherwise, if the start point is located between the target
and the $365^{\circ}$ hard stop, the movement is done in two stages: one from the
start point up to an intermediate point and, secondly, one from this
intermediate point up to the target. The intermediate point is located
between the zero and the target, three motor-steps away from the target.

Although the dispersion of the data points was of the order of a few
microns in all cases, the Single Direction mode gave slightly better
results and therefore was set as the default approach mode for the
actuator. Note that, by comparing the results of ROT1 and ROT2, this
test also shows that the 1:4 gearing stage of ROT1
is hardly affected by any backlash.

\subsubsection{Repeatability tests of ROT1 and ROT2}
%
For the tests here included, some definitions are required
in order to better understand the protocols described next. A target
position is the position which is being measured. A start position is a
position from which the movement to achieve the target position starts.
Therefore, the optical setup is tuned in order to roughly center the target
position on the field of view of the camera, as well as to properly focus
the pinhole spot.
These tests were performed for ROT1 and ROT2 separately. For ROT1, 21
target points were defined, and 11 target points
were defined for ROT2, the difference is due to the smaller range of
ROT2 (see Fig.~\ref{testpoints}). These points give the
reasonable sampling of one target position every 480 steps (for ROT1) and
one target position every 240 half-steps (for ROT2). Each target position
was measured using the rest of the points as starting points (20 start points
for ROT1; 10 start points for ROT2), and with five iterations for each pair start/target
positions. Thus, each target position has produced 100 measurements for ROT1
and 50 for ROT2. This comprehensive series of measurement has, consequently,
produced 2100 measurements for ROT1 and 1100 measurements for ROT2. For each
measurement a picture is captured with the camera.
From each picture, the coordinates of the spot centroid
are found. For each target position, the mean position of the
distribution of centroids (100 items for ROT1; 50 items for ROT2) is found,
as well as the standard deviation ($\sigma$). In addition, the size of the
dispersion box of the distribution is found ($\Delta$). Such data are shown in
Table~\ref{r1performance} for ROT1, and Table~\ref{r2performance} for ROT2.
Since the DoFs of the prototype are
rotational, a polar coordinates system fits better with the present data
process, which means that the size of the dispersion boxes, as well as the
standard deviation, will be given along the angular and the radial
directions.
   \begin{table}
      \caption{ROT1 repeatability performance}
         \label{r1performance}
         \begin{center}       
         \begin{tabular}{|c|c|c|c|c|c|}
            \hline
            \noalign{\smallskip}
            Target & $\sigma_{T}$ & $\sigma_{R}$ & $\Delta_{T}$ & $\Delta_{R}$ & $\sigma_{tot}$\\
            \noalign{\smallskip}
            \hline
            \noalign{\smallskip}
            \rule[-1ex]{0pt}{3.5ex} A  & $3.31$ & $0.31$ & $\pm\,5.42$ & $\pm\,0.72$ & $3.33$\\
            \rule[-1ex]{0pt}{3.5ex} B  & $2.00$ & $1.39$ & $\pm\,3.90$ & $\pm\,2.58$ & $2.44$\\
            \rule[-1ex]{0pt}{3.5ex} C  & $2.41$ & $0.77$ & $\pm\,5.11$ & $\pm\,1.69$ & $2.53$\\
            \rule[-1ex]{0pt}{3.5ex} D  & $2.96$ & $1.44$ & $\pm\,6.33$ & $\pm\,3.93$ & $3.29$\\
            \rule[-1ex]{0pt}{3.5ex} E  & $1.52$ & $2.17$ & $\pm\,3.77$ & $\pm\,5.08$ & $2.65$\\
            \rule[-1ex]{0pt}{3.5ex} F  & $3.12$ & $0.55$ & $\pm\,5.68$ & $\pm\,1.27$ & $3.17$\\
            \rule[-1ex]{0pt}{3.5ex} G  & $2.48$ & $0.38$ & $\pm\,5.64$ & $\pm\,0.70$ & $2.51$\\
            \rule[-1ex]{0pt}{3.5ex} H  & $3.00$ & $0.33$ & $\pm\,5.20$ & $\pm\,0.79$ & $3.02$\\
            \rule[-1ex]{0pt}{3.5ex} I  & $2.59$ & $0.81$ & $\pm\,5.70$ & $\pm\,1.94$ & $2.72$\\
            \rule[-1ex]{0pt}{3.5ex} J  & $2.65$ & $0.41$ & $\pm\,4.64$ & $\pm\,0.81$ & $2.69$\\
            \rule[-1ex]{0pt}{3.5ex} K  & $4.25$ & $1.32$ & $\pm\,8.71$ & $\pm\,2.51$ & $4.45$\\
            \rule[-1ex]{0pt}{3.5ex} L  & $1.89$ & $0.21$ & $\pm\,4.55$ & $\pm\,0.59$ & $1.91$\\
            \rule[-1ex]{0pt}{3.5ex} M  & $2.97$ & $0.33$ & $\pm\,4.63$ & $\pm\,0.67$ & $2.99$\\
            \rule[-1ex]{0pt}{3.5ex} N  & $2.59$ & $0.60$ & $\pm\,5.33$ & $\pm\,1.45$ & $2.65$\\
            \rule[-1ex]{0pt}{3.5ex} P  & $3.38$ & $0.32$ & $\pm\,5.45$ & $\pm\,0.64$ & $3.40$\\
            \rule[-1ex]{0pt}{3.5ex} Q  & $1.07$ & $0.51$ & $\pm\,3.17$ & $\pm\,0.87$ & $1.18$\\
            \rule[-1ex]{0pt}{3.5ex} R  & $2.33$ & $0.37$ & $\pm\,4.02$ & $\pm\,0.91$ & $2.36$\\
            \rule[-1ex]{0pt}{3.5ex} S  & $0.96$ & $0.37$ & $\pm\,2.28$ & $\pm\,0.76$ & $1.03$\\
            \rule[-1ex]{0pt}{3.5ex} T  & $3.32$ & $0.26$ & $\pm\,5.14$ & $\pm\,0.61$ & $3.33$\\
            \rule[-1ex]{0pt}{3.5ex} U  & $1.38$ & $0.22$ & $\pm\,3.61$ & $\pm\,0.51$ & $1.39$\\
            \rule[-1ex]{0pt}{3.5ex} V  & $3.23$ & $0.64$ & $\pm\,5.64$ & $\pm\,2.16$ & $3.30$\\
            \noalign{\smallskip}
            \hline
         \end{tabular}
         \end{center}
   \end{table}
   \begin{table}
      \caption{ROT2 repeatability performance}
         \label{r2performance}
         \begin{center}       
         \begin{tabular}{|c|c|c|c|c|c|}
            \hline
            \noalign{\smallskip}
            Target & $\sigma_{T}$ & $\sigma_{R}$ & $\Delta_{T}$ & $\Delta_{R}$ & $\sigma_{tot}$\\
            \noalign{\smallskip}
            \hline
            \noalign{\smallskip}
            \rule[-1ex]{0pt}{3.5ex} A & $1.52$ & $0.34$ & $\pm\,3.19$ & $\pm\,0.64$ & $1.56$\\
            \rule[-1ex]{0pt}{3.5ex} B & $1.40$ & $0.16$ & $\pm\,2.68$ & $\pm\,0.39$ & $1.41$\\
            \rule[-1ex]{0pt}{3.5ex} C & $1.27$ & $0.44$ & $\pm\,2.36$ & $\pm\,0.68$ & $1.35$\\
            \rule[-1ex]{0pt}{3.5ex} D & $1.75$ & $0.55$ & $\pm\,2.43$ & $\pm\,0.80$ & $1.83$\\
            \rule[-1ex]{0pt}{3.5ex} E & $2.07$ & $0.38$ & $\pm\,2.92$ & $\pm\,0.81$ & $2.10$\\
            \rule[-1ex]{0pt}{3.5ex} F & $1.22$ & $0.19$ & $\pm\,2.44$ & $\pm\,0.34$ & $1.24$\\
            \rule[-1ex]{0pt}{3.5ex} G & $1.01$ & $0.30$ & $\pm\,1.94$ & $\pm\,0.90$ & $1.05$\\
            \rule[-1ex]{0pt}{3.5ex} H & $0.64$ & $0.17$ & $\pm\,0.94$ & $\pm\,0.59$ & $0.67$\\
            \rule[-1ex]{0pt}{3.5ex} I & $0.70$ & $0.18$ & $\pm\,1.48$ & $\pm\,0.57$ & $0.73$\\
            \rule[-1ex]{0pt}{3.5ex} J & $0.51$ & $0.14$ & $\pm\,0.81$ & $\pm\,0.33$ & $0.53$\\
            \rule[-1ex]{0pt}{3.5ex} K & $0.74$ & $0.16$ & $\pm\,1.31$ & $\pm\,0.53$ & $0.76$\\
            \noalign{\smallskip}
            \hline
         \end{tabular}
         \end{center}
   \end{table}
This data process has been implemented for all the target positions. A
further treatment of the data consisted of overlapping the distributions of
all the target points with the mean centroid as a reference. From this
``overall-distribution'', the standard deviation was again found, as well as
the confidence box for $100\%$ of the data.\\
The aim of this process lies on providing some few parameters which may give key
information about the repeatability of the prototype.
The results obtained were as follows:
\begin{itemize}
\item[-] ROT1: Polar components of the standard deviation on $100\%$ of the data
are $\sigma_{T1} = 2.7\mu$ (tangential value) and $\sigma_{R1} = 0.8\mu$ (radial value).\\
$100\%$ of the data fell into a box of $\pm8.7\,\mu$ in tangential direction
and $\pm5.7\,\mu$ in radius.
\item[-] ROT2: Polar components of the standard deviation on $100\%$ of the data
are $\sigma_{T2} = 1.3\mu$ (tangential value) and $\sigma_{R2} = 0.3\mu$ (radial value).\\
$100\%$ of the data fell into a box of $\pm3.6\,\mu$ in tangential direction
and $\pm1.0\,\mu$ in radius.
\end{itemize}
Thus, the prototype can be said to provide an overall repeatability of $\sigma_{TOT} = 3.1\mu$,
the $100\%$ of events being enclosed in a box of $\pm12.3\,\mu$. This performance is well
within the requirements ($40\,\mu$ radius) at $100\%$ level of confidence.

\subsubsection{Soft-zero tests}
There is a ``soft-zero'' position for each of the rotations. A ``soft-zero''
position is the real zero position to which all the angular positions of the
rotation are referred. A software routine has been designed to achieve
this position. Once the rotation has been driven up to the physical stop, it
turns from there until an index mark on the encoder is found. Then, the
motor stops turning at this point.
Through the present test the repeatability of the soft-zero positions of
ROT2 have been measured. The same test on ROT1 was not possible because
of a problem with the encoder of the ROT1 motor.
The dispersion so found will be taken into account for the
results from the accuracy tests. Indeed, the accuracy at certain position is
affected by the repeatability of the soft-zero in the case of power failure.
Since such a failure would imply a reset of the soft-zero positions, a
repeatability error on this position would affect all the positions of that
rotation.
The routine above described was performed 45 times, the overall dispersion box (at
$100\%$ confidence) and the overall standard deviation so obtained being:
$\Delta = \pm5.5\,\mu$, $\sigma = 3.97\,\mu$.


\subsection{Lifetime tests}
Finally, the prototype was submitted to a non-stop run for seven days in
order to check if very long runs could affect its performance. This
accelerated test allowed simulating the entire lifetime of the actuator.
Indeed, each day of run is approximately equivalent to one year of operation
at the telescope. Once again,
in order to set realistic conditions for the present test, a randomised
routine was implemented. The prototype was thus driven continuously through
a sequence of random positions, during seven days. In fact, the present test
was much tougher than normal operation at the telescope because of the
non-stop running. Once this test was finished a few positions were measured
again in order to check whether the performance had changed.
No significant variations were observed as compared to the performance of
the prototype before the lifetime test. 

%
%
%
%
%
%
%
%
%
%
\section{Conclusions and further tasks}
\label{sec:Conclusions}  

The mechanical solution here provided for a fiber positioner actuator fits
by far the requirements in terms of repeatability: $12.3\,\mu$ was
obtained as maximal radius of error ($40\,\mu$ was required).\\
Although this parameter is not directly applicable to the operation conditions
(it is quite unlikely to repeat several times a certain position during
observations at the telescope, within the lifetime of an actuator), it gives
a significant feeling about the good performance of the mechanism.

The procedure of approaching any target points always from the same
direction has been set by default since it gives better results in terms of
repeatability. 

The radial component of the repeatability error (both in ROT1 and ROT2) is
clearly lower than the polar component. This means that the guiding of both
DoFs is extremely accurate.

The worst-case reconfiguration time slightly exceeds the time requirement,
but preliminary simulations show that such worst-case events are quite
rare, thus the requirement should be fulfilled during most of the
operation time.

By using the same design concept, this solution can be extensible to smaller
sizes by a factor down to 0.8.

Thermal issues are involved by default because the results shown here
comprehend any eventual drift produced by temperature variations affecting
the experimental setup and the prototype (motor energy mode is permanently
enabled) during the tests.
Therefore, these results are on the conservative side. Anyway, this issue was
already taken into account when the experimental setup was designed
(the test environment was cooled by a fan and a cold lamp was used).

The main task still pending is the testing of the absolute position error of the
actuator.

The present work has presented a concept for a fiber positioner actuator,
easily scalable for large focal planes, with very high performance in terms
of repeatability. Although the absolute position measurements are still pending
(to be held at IMH in summer 2010 through a 3D coordinate measuring machine)
the concept here presented has successfully passed all the tests done so far.

There is a patent pending for the actuator concept design.

%
%
%
%
%
\acknowledgements
We acknowledge the fundamental contribution of D. Schlegel, B. Ghiorso.
We are grateful to N. Andersen, R. LaFever, R. Wells at the Lawrence Berkeley National Lab.
and M. Colless and S. Barden at the Anglo-Australian Observatory for the PDR review.\\
We are grateful to the Instituto de M\'aquina-Herramienta, Elg\'oibar, Spain (IMH) for
providing the 3D measuring machine.\\
Part of this work was supported by funds from the Spanish ``Centro para
el desarrollo Tecnol\'ogico Industrial'' (CDTI) and the
Spanish ``Ministerio de Ciencia e Innovaci\'on'' (MICINN).


%
%

\end{document}